%% file: Template.tex
\documentclass{article}
\usepackage{spconf,amsmath,graphicx, subfig, lipsum}
\usepackage{pgfplots}
\usepackage{tikz}
\usepackage{bm}
\usepackage{amsfonts}
\usepackage{mathtools}
\usepackage{dblfloatfix}

\newcommand{\floor}[1]{\lfloor #1 \rfloor}

\title{Learned Image Compression with Soft Bit-based Rate-Distortion Optimization}

\input{"chapter/header.tex"}

\begin{document}
\maketitle

% Abstract.
%------------------------------------------------------------------------
\input{"chapter/abstract.tex"}

% 1. Introduction.
%------------------------------------------------------------------------
\input{"chapter/1-introduction.tex"}

% 2. Methods.
%------------------------------------------------------------------------
\input{"chapter/2-method.tex"}

% 3. Training.
%------------------------------------------------------------------------
\input{"chapter/3-training.tex"}
%------------------------------------------------------------------------

% 4. Evaluation.
%------------------------------------------------------------------------
\input{"chapter/4-evaluation.tex"}
%------------------------------------------------------------------------

% 5. Conclusion.
%------------------------------------------------------------------------
\input{"chapter/5-conclusion.tex"}
%------------------------------------------------------------------------

% 6. Acknowledgement.
%------------------------------------------------------------------------
%\input{"chapter/6-acknowledgement.tex"}
%------------------------------------------------------------------------

% 7. References.
%------------------------------------------------------------------------
\newpage
\bibliographystyle{IEEEbib}
\bibliography{refs}
%------------------------------------------------------------------------

\end{document}

%% file: chapter/header.tex
\name{David Alexandre \qquad Chih-Peng Chang \qquad Wen-Hsiao Peng \qquad Hsueh-Ming Hang}
\address{National Chiao Tung University, Taiwan}

%% file: chapter/abstract.tex
\begin{abstract}
This paper introduces the notion of soft bits to address the rate-distortion optimization for learning-based image compression. Recent methods for such compression train an autoencoder end-to-end with an objective to strike a balance between distortion and rate. They are faced with the zero gradient issue due to quantization and the difficulty of estimating the rate accurately. Inspired by soft quantization, we represent quantization indices of feature maps with differentiable soft bits. This allows us to couple tightly the rate estimation with context-adaptive binary arithmetic coding. It also provides a differentiable distortion objective function. Experimental results show that our approach achieves the state-of-the-art compression performance among the learning-based schemes in terms of MS-SSIM and PSNR.

\end{abstract}
\begin{keywords}
Autoencoder, Deep Learning,  Image Compression, Soft Bits
\end{keywords}

%% file: chapter/1-introduction.tex
\section{Introduction}
\label{sec:intro}

%background
Learning-based image compression has recently attracted lots of attention due to the renaissance of deep learning. Unlike the traditional methods, the learning-based schemes can be adapted to any differentiable objective, opening up many optimization possibilities. For example, Li \emph{et al.}~\cite{li2018learning} propose a content-weighted image compression model that performs region-adaptive compression via a learnable importance map.

%problems
Most learning-based methods\cite{li2018learning, rippel2017real, mentzer2018conditional, minnen2018joint, agustsson2017soft, balle2016end, balle2018variational, theis2017lossy} rely on training an autoencoder end-to-end with the aim of striking a good balance between distortion and rate losses. Two challenges arise. First, the quantization process for lossy feature map compression causes zero gradients during the back-propagation process. Second, the rate loss is often painful to estimate accurately, as it is highly coupled with entropy coding, the operation of which is generally not differentiable.   

%quick survey
Several prior arts are proposed to address these issues. Li \emph{et al.}~\cite{li2018learning} overcome the zero gradients by a straight-through mechanism, which simply considers the quantizer to be an identity function during the back-propagation process. Agustsson \emph{et al.}~\cite{agustsson2017soft} and Mentzer \emph{et al.}~\cite{mentzer2018conditional} introduce a non-uniform soft quantizer with a smooth mapping function as a surrogate of the hard quantizer. Ball{\'e} \emph{et al.}~\cite{balle2016end, balle2018variational} and Theis \emph{et al.}~\cite{theis2017lossy}  adopt an additive noise model for the quantizer. 

%quick survey-2
In comparison with the quantization issue, the rate estimation is even more challenging. Li \emph{et al.}~\cite{li2018learning} use the sum of importance map features as a rough estimate of the rate. Theis \emph{et al.}~\cite{theis2017lossy} estimate the rate from the upper-bound of non-differentiable number of bits. For better estimation, Ball{\'e} \emph{et al.}~\cite{balle2016end, balle2018variational} and Minnen \emph{et al.}~\cite{minnen2018joint} compute the differential entropy of the quantizer output based on the additive noise model. To bind the rate estimation tightly to the actual entropy coding, Mentzer \emph{et al.}~\cite{mentzer2018conditional} use the context probability model implemented by PixelRNN \cite{van2016pixel} to compute the self-information of each coding symbol. Their scheme is, however, complicated due to the use of PixelRNN \cite{van2016pixel} and the non-binary arithmetic coding.

%how to solve the problem compared to others
In this paper, we propose a learned image compression system with soft-bit-based rate-distortion optimization. It has the striking feature of combining effective coding tools from modern image codecs (e.g., uniform quantization, binary bitplane coding with on-the-fly probability updating, and simple context models) with the strong suit of deep learning (e.g., non-linear autoencoder). Moreover, we introduce the notion of soft bits to represent quantization indices of feature samples so that both rate and distortion losses can be estimated accurately in a differentiable manner. Experimental results show that our method achieves the state-of-the-art rate-distortion performance among the learning-based schemes.  

%section details.
The remainder of this paper is organized as follows: Section 2 describes the proposed method. Section 3 details the training procedure. Section 4 presents the experimental results. Section 5 concludes this work.

%% file: chapter/2-method.tex
\begin{figure*}
\centering
        \resizebox{1\textwidth}{!}{
        \begin{tikzpicture}[node distance=4cm]
        %\draw [help lines] (0,0) grid (30,2);
        \tikzstyle{entity_blue}=[minimum height=1cm,minimum width=3cm, draw=blue!50,fill=blue!20,thick]
        %\tikzstyle{entity_gray}=[minimum height=1cm,minimum width=3cm, draw=gray!50,fill=gray!20,thick]
        \tikzstyle{entity}=[minimum height=1cm,minimum width=3cm, draw=black,thick]
        \node                 (x)   at (-1,0) {\LARGE$\bm{x}$};  
        \node[entity_blue]    (e)   at (2,0) {\Large Encoder $\theta_e$};
        \node[entity]         (q)   [right of=e] {\Large Q}; 
        \node[entity]         (sq)   at (6,-2) {\Large SB Conv.}; 
        \node[entity]         (ac)  [right of=q] {\Large CABIC}; 
        \node[entity_blue]         (r)   at (14,-3) {\Large Rate Estimator $\theta_r$};
        \node                      (D)   at (14,1.5) {\Large $L_D(\bm{x},\hat{\bm{x}})$};
        \node                 (bit) [right of=ac] {\Large Bitstream};
        \node[entity]         (dac) [right of=bit] {\Large CABID};  
        \node[entity]         (iq)  [right of=dac] {\Large IQ};
        \node[entity]         (isq)  at (22,-2) {\Large Inv. SB Conv.};
        \node[entity_blue]         (d)   [right of=iq] {\Large Decoder $\theta_d$};
        \node                 (x_hat)  at (29,0) {\LARGE$\hat{\bm{x}}$};
        \foreach \from/\to in {x/e, ac/bit, bit/dac, iq/d,d/x_hat}
           \draw [->,line width=1.5pt] (\from) -- (\to);
        \draw [->,line width=1.5pt] (e) -- (q) node (f) [midway, above]{\LARGE$\bm{f}$};
        \draw [->,line width=1.5pt] (q) -- (ac) node [midway, above]{\LARGE$\bm{q}$};
        \draw [->,line width=1.5pt] (dac) -- (iq) node [midway, above]{\LARGE$\bm{q}$};
        \draw [->,line width=1.5pt] (iq) -- (d) node (f_hat) [midway, above]{\LARGE$\hat{\bm{f}}$};
        \draw[->,line width=1.5pt, dashed] (0,0) |- (D);
        \draw[->,line width=1.5pt, dashed] (28,0) |- (D);
        \draw [->,line width=1.5pt, dashed] (sq) -- (isq) node (q_tilde) [midway, above] {\LARGE$\tilde{\bm{q}}$};
        \draw [->,line width=1.5pt, dashed] (f) |- (sq);
        \draw [->,line width=1.5pt, dashed] (isq) -| (f_hat);
        \draw [->,line width=1.5pt, dashed] (q_tilde) -- (r);
        \end{tikzpicture}
        }
\caption{The architecture of the proposed image compression model.}
\label{fig:network}
\vspace{-0.3cm}
\end{figure*}
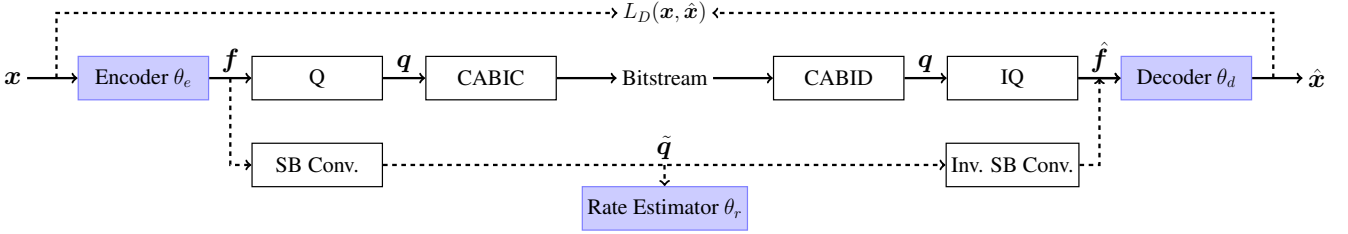
\section{Proposed Method}
\label{sec:system_designs}
This section details the framework of our image compression system, including the overall architecture, the operation of each component, and the modeling of compression rate and distortion for end-to-end training. Notation-wise, we use a bold letter (e.g., $\bm{x}$) to refer collectively to a high-dimensional tensor and a Roman letter (e.g., $x$) to denote its element in some order.  

%2.1
\subsection{Overall Architecture}
Fig. \ref{fig:network} illustrates our proposed framework. There are two data paths, one for operating the model in the test mode (that is, for putting it into use in practice) and the other for its training (i.e., training mode). 

The data path in the test mode, as indicated by the solid arrow lines, begins with encoding an image $\bm{x} \in \mathbb{R}^{W \times H \times 3}$ of size $W \times H$ in 4:4:4 YUV format through a convolutional encoder $E(\bm{x};\theta_e)$ into a compact set of feature maps $\bm{f} \in \mathbb{R}^{W/8 \times H/8 \times C}$, of which each feature sample $f \in (0,1)$ is a real number. For lossy compression, $f$ is uniformly quantized by a $b$-bit, power-of-two quantizer $Q$, leading to a fixed-point binary representation $q=\floor{f/2^{-b}}$, where $2^{-b}$ is the quantization step size. That is,  the quantization (output) index $q$ is the first $b$ significant bits of $f$ in its binary representation (e.g., $q=1100$ for $f=0.81,b=4$). Like most image compression systems, either learning-based or conventional, the quantization indices $\bm{q}$ are compacted further by lossless arithmetic coding. Motivated by JPEG2000 \cite{taubman2012jpeg2000}, we arrange $\bm{q}$ as bitplanes and perform context-adaptive bit-plane encoding/decoding (CABIC/CABID), of which we will discuss more in the following sections. To reconstruct the input $\bm{x}$ approximately, the feature sample is first recovered via inverse quantization (IQ) $\hat{f}=q/2^b$, followed by convolutional decoding $\hat{\bm{x}}=D(\hat{\bm{f}};\theta_d)$. Currently, our encoder and decoder come from an autoencoder proposed in \cite{mentzer2018conditional}; their parameters $\theta_e,\theta_d$ are however learned by our training framework, which aims to strike a good trade-off between rate $L_R(\bm{q})$ and distortion $L_D(\bm{x},\hat{\bm{x}})$ by minimizing the following objective function with respect to $\theta_e,\theta_d$:
\begin{equation}
    \lambda \times L_R(\bm{q}) + L_D(\bm{x},\hat{\bm{x}}),
\end{equation}
where $L_D(\bm{x},\hat{\bm{x}})$ is defined to be a weighted sum of mean-square errors between YUV components of $\bm{x}$ and $\hat{\bm{x}}$, with the error of Y component weighted 4 times that of the U/ V component. 

The data path in training mode, as outlined by the dashed arrow lines, is designed for end-to-end model training. Training a learning-based compression system is often faced with two issues: (1) the quantization effect, which describes the stair-like mapping from $\bm{f}$ to $\hat{\bm{f}}$, gives rise to zero gradients almost everywhere, and (2) the rate cost needed to achieve a rate-distortion optimized design is difficult to estimate accurately. To address these issues, we introduce the notion of \textit{soft bits} $\tilde{\bm{q}}$ as an alternative to the \textit{hard bit} representation of the quantization indices $\bm{q}$. As an example, instead of rendering $q$ into "1","1","0","0" for $f=0.81,b=4$ as done previously, we express these binary hard bits as real-valued soft bits, e.g. "0.91", "0.95", "0.1", "0.07", by the soft bit conversion (SB Conv.) module. In doing so, each of these soft bits is formulated as a differentiable function of $f$. Not only can they be used together with a differentiable rate estimator, implemented by a learnable neural network with parameter $\theta_r$ in Fig. \ref{fig:network}, to give an accurate estimate of the coding cost, but they can also be used to approximate $\hat{\bm{f}}$ in a differentiable manner (by the Inv. SB Conv. module). 

To sum up, our framework has three networks to be learned end-to-end: the encoder, the decoder, and the rate estimator. Among these, only the encoder and the decoder will actually operate in the test mode, while the rate estimator is activated for training only.      

%2.2
\subsection{Soft Bit Conversion}
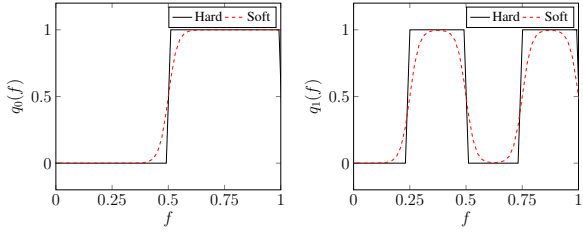
\begin{figure}[!t]
\centering
        \resizebox{0.45\linewidth}{!}{
        \begin{tikzpicture}[
                declare function={ 
                        Sigmoid(\x) = 1/(1+exp(-50*\x));
                }]
        \begin{axis}[
        name=plot1,
        xlabel={\Large$f$},
        xmin=0,xmax=1,
        xtick={0,0.25,...,1},
        ylabel={\Large$q_0(f)$},
        ymin=-0.2,ymax=1.2,
        samples=500,
        legend columns=2,
        ticklabel style = {font=\Large},
        legend style={font=\large}
        ]
        \addplot[mark=none] {round(mod(x,1))};
        \addplot[mark=none,red, dashed] {Sigmoid(x-0.5)};
        \legend{Hard,Soft};
        \end{axis}
        \end{tikzpicture}
        }
        \resizebox{0.45\linewidth}{!}{
        \begin{tikzpicture}[
                declare function={ 
                        Sigmoid(\x) = 1/(1+exp(-50*\x));
                }]
        \begin{axis}[
        name=plot2,
        xlabel={\Large$f$},
        xmin=0,xmax=1,
        xtick={0,0.25,...,1},
        ylabel={\Large$q_1(f)$},
        ymin=-0.2,ymax=1.2,
        samples=500,
        legend columns=2,
        ticklabel style = {font=\Large},
        legend style={font=\large}
        ]
        \addplot[mark=none] {round(mod(2*x,1))};
        \addplot[mark=none,red, dashed] {Sigmoid(x-0.25)-Sigmoid(x-0.5)+Sigmoid(x-0.75)-Sigmoid(x-1.0)};
        \legend{Hard,Soft};
        \end{axis}
        \end{tikzpicture}
        }
%\vspace{-0.2cm}
\caption{Soft bit versus hard bit mappings.}
\label{fig:soft_bit_generation}
%\vspace{-0.4cm}
\end{figure}

The soft bit conversion plays a central role in enabling our compression system end-to-end trainable. It is to convert the binary, hard-bit representation of the quantization index $q$ of a feature sample $f$ into a differentiable function of $f$, namely the soft-bit representation. In the previous example,  the binary fixed-point representation of $q$ for a feature sample $f=0.81$ is "1100" when $f$ is quantized uniformly with a step size of $2^{-4}$. We observe that each of these hard bits $q_0=1$, $q_1=1$, $q_2=0$, $q_3=0$ is in fact a function of $f$. For instance, the first bit $q_0$ equals to 1 when $f$ is in the interval $[0.5,1)$ and 0 when in the interval of $[0,0.5)$. The mappings for the first two bits $q_0,q_1$ are visualized in Fig. \ref{fig:soft_bit_generation} (see the hard-bit curves). Apparently, due to their rectangular waveforms, the derivative with respect to $f$ is zero almost everywhere, making the training with back-propagation impossible.

To circumvent this difficulty, we approximate these hard-bit mappings by a superposition of sigmoid functions (see the soft-bit curves in Fig. \ref{fig:soft_bit_generation}). This is motivated by the fact that any rectangular waveform can be expressed as a superposition of step functions, which in turn can be approximated by sigmoid functions with a adequately chosen hyper-parameter $\alpha$:
\begin{equation}
    u(f) \coloneqq 
    \left\{
	    \begin{array}{ll}
		    1  & \mbox{if } f \geq 0 \\
		    0 & \mbox{if } f < 0
	    \end{array}
    \right.
    \approx 
    \sigma_{\alpha}(f) \coloneqq  \frac{1}{1+e^{-\alpha f}}.
\end{equation}
As an example, it is seen that:
\begin{align}
    q_1(f) & \approx \tilde{q}_1(f) \nonumber
    \\ & \coloneqq 
    \sigma_{\alpha}(f-0.25) -      \sigma_{\alpha}(f-0.5) \nonumber
    \\
    & + \sigma_{\alpha}(f-0.75) -   \sigma_{\alpha}(f-1).
\end{align}
With this approximation, $\hat{f}$ is modeled by the soft bits using $\tilde{q}_0 \times 2^{-1} + \tilde{q}_1 \times 2^{-2} + \tilde{q}_2 \times 2^{-3} + \tilde{q}_3 \times 2^{-4}$ in the back-propagation process. Note that one may as well use the soft quantization technique in \cite{mentzer2018conditional} to model the mapping from $\bm{f}$ to $\hat{\bm{f}}$ directly. 

Although our current model implements a power-of-two uniform quantizer, the soft-bit representation for quantization indices can readily be applied to non-uniform quantizers.

%2.4 
\subsection{Context-adaptive Bit-plane Coding (CABIC)}
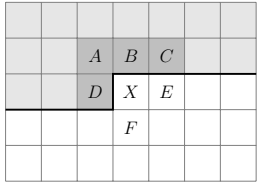
\begin{figure}
\centering
        \resizebox{0.4\linewidth}{!}{
        \begin{tikzpicture}
        \fill[fill=gray!20] (0,2) -- (3,2) -- (3,3) -- (7,3) -- (7,5) -- (0,5) -- cycle;
        \fill[fill=gray!50] (2,2) -- (3,2) -- (3,3) -- (5,3) -- (5,4) -- (2,4) -- cycle;
        
        \draw[color=gray] (0,0) grid (7,5);
        
        \pgfsetlinewidth{1.5pt};
        \pgfpathmoveto{\pgfpointxy{0}{2}};
        \pgfpathlineto{\pgfpointxy{3}{2}};
        \pgfpathlineto{\pgfpointxy{3}{3}};
        \pgfpathlineto{\pgfpointxy{7}{3}};
        \pgfusepath{stroke};
        
        \node at (3.5, 2.5) {\Large $X$};
        \node at (2.5, 3.5) {\Large $A$};
        \node at (3.5, 3.5) {\Large $B$};
        \node at (4.5, 3.5) {\Large $C$};
        \node at (2.5, 2.5) {\Large $D$};
        \node at (4.5, 2.5) {\Large $E$};
        \node at (3.5, 1.5) {\Large $F$};
        \end{tikzpicture}
        }
%\vspace{-0.2cm}
\caption{Illustration of context template for context-adaptive bit-plane coding. $X$ denotes the location of a coding sample.}
\label{fig:ctx_design}
\vspace{-0.4cm}
\end{figure}

Before describing our soft-bit-based rate estimation, we present briefly how the quantization indices $\bm{q}$ of feature maps $\bm{f}$ are coded in the test mode. We first organize $\bm{q}$ into bitplanes. A bitplane is formed collectively by the same binary digits of quantization indices. For example, the most significant bitplane consists of all the $q_0$ of feature samples. Bits are then coded starting from the most significant bitplane to the least significant one, with different feature maps processed in the same manner yet separately. 

To encode a bitplane, we adopt the context-adaptive binary arithmetic coding technique. Inspired by JPEG2000, we classify every bit into a significant bit or a refinement bit. Using Fig. \ref{fig:ctx_design} for illustration, for coding a significant bit of the quantization index at $X$, we refer to the binary significant status of the surrounding indices at $B$, $D$, $E$ and $F$. This yields a total of 16 context patterns (or ctx values for short), each corresponding to a binary probability model that is updated on-the-fly. For coding a refinement bit, the ctx value is computed based on the bit values of quantization indices at $B,D,E,F$ in the previous bitplane along with those of $A,B,C,D$ in the current bitplane. Since refinement bits are less predictable, we reduce the number of their ctx values to 9 only. 

Note that we adopt the traditional hand-crafted design for arithmetic coding because (1) it allows simple adaptation of the context probability model to learn local image statistics and (2) it avoids the need to perform neural network inference at bit level, which introduces extra processing latency in the highly sequential arithmetic decoding process.  

%2.3
\subsection{Rate Estimator}

\begin{figure}
\centering
        \resizebox{0.7\linewidth}{!}{
        \begin{tikzpicture}[node distance=4cm]
        %\draw [help lines] (0,-1) grid (10,1);
        \tikzstyle{entity_blue}=[minimum height=1cm,minimum width=2cm, draw=blue!50,fill=blue!20,thick]
        \tikzstyle{entity}=[minimum height=2.5cm,minimum width=2.5cm, draw=black,thick]
        \node                 (ctx)   at (1,0.3)  {$ctx$};
        \node                 (q)    at (1,-0.3) {$\tilde{q}_i$};
        \node[entity_blue, align=center]    (pr)    at (3,0)  {Probability \\ Regressor \\ $\theta_r$};
        \node                 (p)     at (6,1.2)  {$p({q}_i|ctx)$};
        \node [circle,draw]  (sub)     at (6,0)  {-};         
        %\node                 (o)     at (8,0)  {$-\log p(\tilde{q}_i|ctx)$};
        \node                 (o)     at (7,0)  {};
        \draw [->,line width=1pt] (ctx) -- (2,0.3);
        \draw [->,line width=1pt] (q) -- (2,-0.3);
        \draw [->,line width=1pt] (pr) -- (sub) node [midway, above] {$p(\tilde{q}_i|ctx)$};
        \draw [->,line width=1pt] (p) -- (sub);
        \draw [->,line width=1pt] (sub) -- (o);
        %\draw [dashed,line width=1pt] (1.5,-1.2) rectangle (6.5,1.2);
        %\node                 (r)   at (4,-1.4)  {Rate Estimator}; 
        \end{tikzpicture}        
        }
%\vspace{-0.1cm}
\caption{Training of our probability regressor.}
\label{fig:rate_estimator}
\vspace{-0.4cm}
\end{figure}
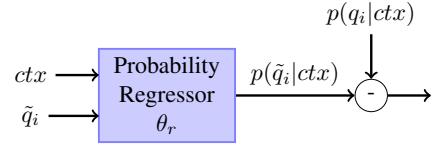

To estimate the code length needed to represent an input bit at training time, we refer to its \textit{self-information}. The self-information of a probabilistic event $\mathcal{E}$ is defined to be the negative logarithm $-\log p(\mathcal{E})$ of its probability $p(\mathcal{E})$. In our case, the probability of a coding bit $q_i$ is maintained in a context probability model, which keeps track of $p(q_i|ctx)$, where $ctx$ denotes its context pattern/value. It is however noted that $p(q_i|ctx)$ is approximated by the relative frequency of $q_i$ given the $ctx$, e.g. how many times the event $q_i=1$ occurs given the present $ctx$, which is a statistics quantity not differentiable with respective to $q_i$.

To overcome this problem, we train a rate estimator that includes a neural network as a probability regressor to fit $p(q_i|ctx)$ collected from the training data, as illustrated in Fig. \ref{fig:rate_estimator}. In particular, the probability regressor takes as input the soft bits version $\tilde{q}_i$ of $q_i$ so that it generates non-zero gradient of the estimated rate (computed to be $-\log p(\tilde{q}_i|ctx)$) with respect to the encoder parameter $\theta_e$:
\begin{equation}
\begin{split}
    \nabla_{\theta_e}(-\log p(\tilde{q}_i|ctx)) = 
    - \frac{1}{p(\tilde{q}_i|ctx)} \frac{\partial p(\tilde{q}_i|ctx)}{\partial \tilde{q}_i} \frac{d \tilde{q}_i }{df} \nabla_{\theta_e} f. 
    \label{eq:rate_est_gradient}
\end{split}    
\end{equation}
It can be seen that if the hard bit mapping is used, the term $d \tilde{q}_i/d f$ would be replaced with $d q_i/d f$, which vanishes.

Eq. \eqref{eq:rate_est_gradient} additionally gives us some important insights into how the estimated rate cost of an input bit $q_i$ would influence the update of the encoder parameter $\theta_e$. Its contribution to the change of $\theta_e$ in a gradient update step will be more significant if $q_i$ is in its less probable state, i.e., $p(\tilde{q}_i|ctx) \leq 0.5$, or if its conditional probability distribution $p(\tilde{q_i}|ctx)$ is more biased, i.e., $\partial p(\tilde{q}_i|ctx)/\partial \tilde{q}_i$ is larger. The latter occurs when $p(\tilde{q_i}=1|ctx) \gg p(\tilde{q_i}=0|ctx)$ or vice versa. 

%% file: chapter/3-training.tex
\section{Training}
\label{sec:training}

The encoder, decoder, and rate estimator are trained in two alternating phases. In the first phase, we collect the statistics of the context probabilities $p(q_i|ctx)$ from the feature maps, and update the rate estimator $\theta_r$ by minimizing the regression error between $p(q_i|ctx)$ and $p(\tilde{q}_i|ctx)$. In the second phase, we incorporate the rate estimator to give an estimate of the rate cost $L_R(\bm{q})$ and update both the encoder and decoder by minimizing $\lambda \times L_R(\bm{q}) + L_D(\bm{x},\hat{\bm{x}})$ with respect to their network parameters $\theta_e,\theta_d$. During training, we set the batch size to 8 and the learning rate to $1e^{-4}$.  

The training dataset contains 1,672 images provided by the Challenge on Learned Image Compression (CLIC) 2018 \cite{CLIC2018}. They are randomly cropped into 128x128 patches, and the horizontal and vertical flipping is performed for data augmentation. 

%% file: chapter/4-evaluation.tex
\section{Experimental results}
\label{sec:evaluation}

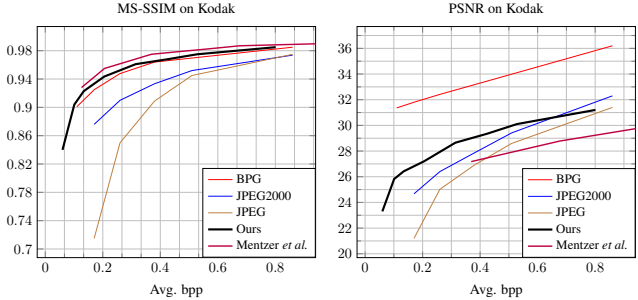
\begin{figure}
\vspace{-0.4cm}
\subfloat{
\begin{tikzpicture}[scale=0.54]
	\begin{axis}[
		xlabel=Avg. bpp,
		%ylabel=MS-SSIM,
		title=MS-SSIM on Kodak,
		grid=both, minor tick num=2,
		ytick={0.7, 0.74,...,1},
		xmax=0.94,
		legend style={at={(axis cs:0.94,0.688)},anchor=south east, font=\small},
		legend cell align={left}
		]
	
	\addplot[color=red] coordinates {
		(0.11, 0.901)
		(0.17, 0.9249428571)
		(0.26, 0.9477506494)
		(0.38, 0.964)
		(0.51, 0.9696875)
		(0.86, 0.985)
	};
	\addplot[color=blue] coordinates {
		(0.17, 0.876)
		(0.26, 0.91)
		(0.38, 0.9333673469)
		(0.51, 0.952)
		(0.86, 0.9738)
	};
	\addplot[color=brown] coordinates {
		(0.17, 0.715)
		(0.26, 0.85)
		(0.38, 0.9089387755)
		(0.51, 0.945)
		(0.86, 0.975)
	};
	\addplot[color=black, line width=1.5pt] coordinates {
		(0.06, 0.84)
		(0.101, 0.904)
		(0.134, 0.923)
		(0.205, 0.9435)
		(0.314, 0.961)
		(0.425, 0.968)
		(0.528, 0.975)
		(0.801, 0.985)
	};
	\addplot[color=purple, line width=1pt] coordinates {
	    (0.126, 0.928)
	    (0.206, 0.955)
		(0.369, 0.975)
		(0.471, 0.979)
		(0.551, 0.982)
		(0.676, 0.987)
		(1.050, 0.991)
	};
	\legend{BPG, JPEG2000, JPEG, Ours, Mentzer \emph{et al.}};
	\end{axis}
\end{tikzpicture}
}
\hfill
\subfloat{
\begin{tikzpicture}[scale=0.54]
	\begin{axis}[
		xlabel=Avg. bpp,
		%ylabel=PSNR,
		title=PSNR on Kodak,
		grid=both, minor tick num=1,
		ytick={20, 22,...,36},
		xmax=0.94,
		legend style={at={(axis cs:0.94,19.7)},anchor=south east, font=\small},
		legend cell align={left},
		]
	
	\addplot[color=red] coordinates {
		(0.11, 31.36)
		(0.17, 31.80307692)
		(0.26, 32.41506494)
		(0.38, 33.16938776)
		(0.51, 33.99017857)
		(0.86, 36.2)
	};
	\addplot[color=blue] coordinates {
		(0.17, 24.67)
		(0.26, 26.4)
		(0.38, 27.85816327)
		(0.51, 29.42)
		(0.86, 32.3)
	};
	\addplot[color=brown] coordinates {
		(0.17, 21.2)
		(0.26, 25.02)
		(0.38, 26.91020408)
		(0.51, 28.6)
		(0.86, 31.4)
	};
	\addplot[color=black, line width=1.5pt] coordinates {
		(0.06, 23.30163725)
		(0.101, 25.82363229)
		(0.134, 26.41953117)
		(0.205, 27.202646)
		(0.314, 28.65294433)
		(0.425, 29.35554692)
		(0.528, 30.10238371)
		(0.801, 31.20489204)
	};
	\addplot[color=purple, line width=1pt] coordinates {
		(0.369, 27.179)
		(0.676, 28.775)
		(1.050, 30.147)
	};
	\legend{BPG, JPEG2000, JPEG, Ours, Mentzer \emph{et al.}};
	\end{axis}
\end{tikzpicture}
}
%\vspace{-0.2cm}
\caption{Rate-distortion comparison on Kodak dataset.}
\label{fig:RD_curve}
%\vspace{-0.4cm}
\end{figure}

\begin{figure}
\centering
\resizebox{\columnwidth}{!}
{
\setlength{\tabcolsep}{1pt}
\begin{tabular}{cc}
\includegraphics[width=0.5 \columnwidth]{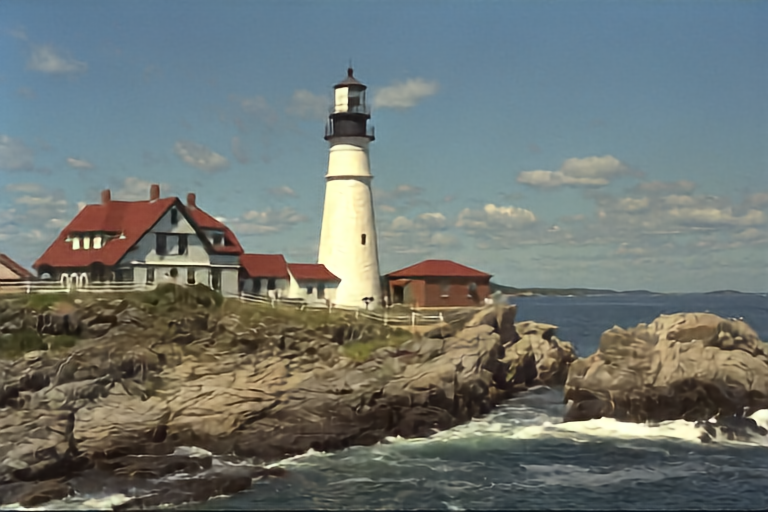}&
\includegraphics[width=0.5 \columnwidth]{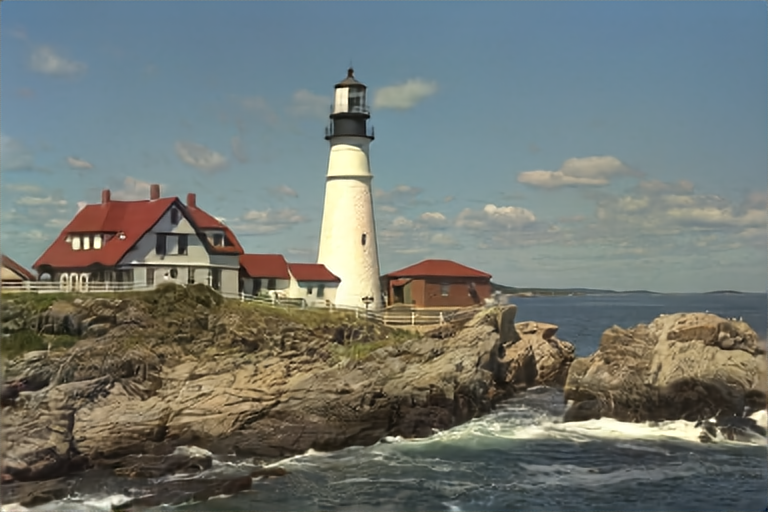}\\
Ours & Mentzer \emph{et al.} \cite{mentzer2018conditional} \\
\includegraphics[width=0.5 \columnwidth]{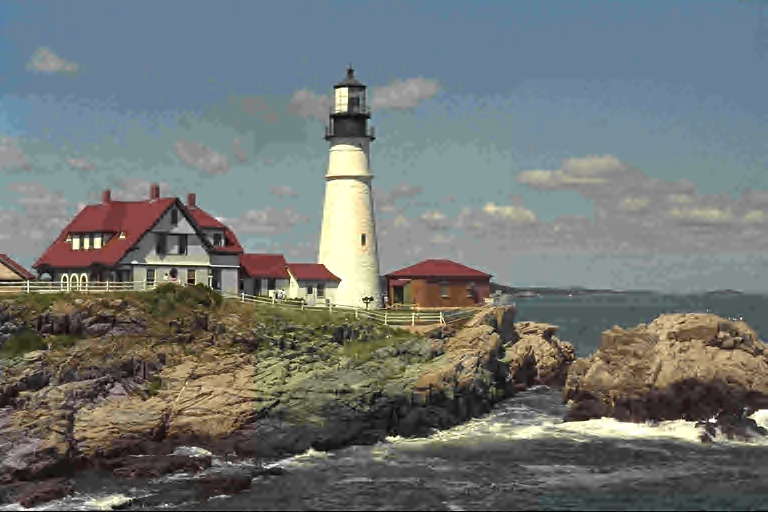}&
\includegraphics[width=0.5 \columnwidth]{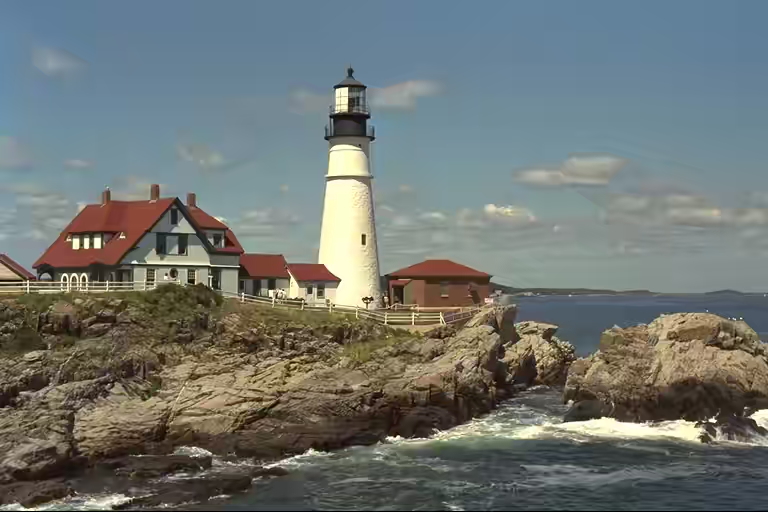}\\
JPEG2000 & BPG\\
\end{tabular}
}
%\vspace{-0.1cm}
\caption{Subjective quality comparison at 0.36 bpp.}
\label{fig:comparison}
\vspace{-0.25cm}
\end{figure} 

\begin{figure}
\centering
\resizebox{\columnwidth}{!}
{
\setlength{\tabcolsep}{1pt}
\begin{tabular}{cc}

\includegraphics[width=0.49 \columnwidth]{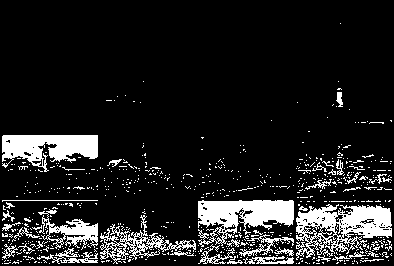}&
\includegraphics[width=0.49 \columnwidth]{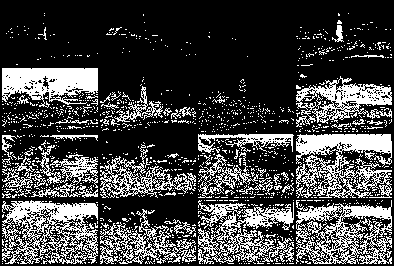}\\
0.06 bpp & 0.12 bpp
\\
\end{tabular}
}
%\vspace{-0.1cm}
\caption{ Visualization of our feature maps at different rates. Each rate displays four bitplanes of one feature map per column.}

\label{fig:featuremap-bpp-comparison}
\vspace{-0.45cm}
\end{figure}

This section compares the rate-distortion performances of the proposed method with the other codecs.  The comparison is conducted on Kodak dataset \cite{kodak} by compressing test images at several rates with a varying number of feature maps. Specifically, our encoder is configured to produce 4 feature maps for bits-per-pixel (bpp) lower than 0.25, 8 for bpp's between 0.25 and 0.5, and 16 for bpp's higher than 0.5. For every test image, we first calculate the average PSNR and MS-SSIM over its three color components. We then present the average values over the entire dataset as a single quality indicator.

From Fig. \ref{fig:RD_curve},  we see that our method performs comparably to BPG and Mentzer \emph{et al.}'s \cite{mentzer2018conditional} while outperforming JPEG and JPEG2000 by a large margin across a wide range of bpp's. On the other hand, in terms of PSNR, it is much inferior to BPG but is superior to the other baselines. These observations are in line with the findings of the other researchers that the learning-based methods often show much better MS-SSIM performance, especially at low rates. It is worth pointing out that our model is trained by minimizing the mean-squared error while Mentzer \emph{et al.}~\cite{mentzer2018conditional} optimize theirs for MS-SSIM. This explains why their method has low PSNR. Fig. \ref{fig:comparison} further displays reconstructed images produced by these codecs for subjective quality evaluation. 

Fig. \ref{fig:featuremap-bpp-comparison} shows the bit allocation among feature maps due to our soft-bit-based rate-distortion optimization. Three observations can be made: (1) the dynamic range of feature samples is adjusted by the encoder depending on the compression rate, as evidenced by the zero bitplanes at lower bpp's; (2) some feature maps are more important than the others in the rate-distortion sense, as evidenced by the uneven bit distribution across feature maps; and (3) the bit allocation is spatially varying, as indicated by the uneven bit distribution across different regions. These together produce a net effect similar in spirit to the importance map mechanism \cite{li2018learning}.

%From Fig. \ref{fig:featuremap-bpp-comparison}, we can observe some interesting 

%We can see that our context model influences the system to recognize spatial regions across feature maps and distribute the bits to meet the target. 

%% file: chapter/5-conclusion.tex
\section{Conclusion}
\label{sec:conclusion}
This paper introduces a learned image  compression system with soft-bit-based rate-distortion optimization. The soft bit representation allows the rate estimation to be tightly coupled with entropy coding, giving an accurate rate estimate. We also show that learning-based compression methods can leverage well-designed coding tools from modern image codecs for a more cost-effective solution. 
%The future works can explore different mapping functions to enhance the gradient approximation and exploring more powerful context model to gain better performance. This may lead to better compression performance and further reducing the approximation error.

%% file: Template.bbl
\begin{thebibliography}{10}

\bibitem{li2018learning}
M.~Li, W.~Zuo, S.~Gu, D.~Zhao, and D.~Zhang,
\newblock ``Learning convolutional networks for content-weighted image
  compression,''
\newblock in {\em Proceedings of the IEEE Conference on Computer Vision and
  Pattern Recognition}, 2018, pp. 3214--3223.

\bibitem{rippel2017real}
O.~Rippel and L.~Bourdev,
\newblock ``Real-time adaptive image compression,''
\newblock in {\em International Conference on Machine Learning}, 2017, pp.
  2922--2930.

\bibitem{mentzer2018conditional}
F.~Mentzer, E.~Agustsson, M.~Tschannen, R.~Timofte, and L.~Van~Gool,
\newblock ``Conditional probability models for deep image compression,''
\newblock in {\em IEEE Conference on Computer Vision and Pattern Recognition
  (CVPR)}, 2018, vol.~1, p.~3.

\bibitem{minnen2018joint}
D.~Minnen, J.~Ball{\'e}, and G.~D. Toderici,
\newblock ``Joint autoregressive and hierarchical priors for learned image
  compression,''
\newblock in {\em Advances in Neural Information Processing Systems}, 2018, pp.
  10794--10803.

\bibitem{agustsson2017soft}
E.~Agustsson, F.~Mentzer, M.~Tschannen, L.~Cavigelli, R.~Timofte, L.~Benini,
  and L.V. Gool,
\newblock ``Soft-to-hard vector quantization for end-to-end learning
  compressible representations,''
\newblock in {\em Advances in Neural Information Processing Systems}, 2017, pp.
  1141--1151.

\bibitem{balle2016end}
J.~Ball{\'e}, V.~Laparra, and E.~P Simoncelli,
\newblock ``End-to-end optimized image compression,''
\newblock {\em arXiv preprint arXiv:1611.01704}, 2016.

\bibitem{balle2018variational}
J.~Ball{\'e}, D.~Minnen, S.~Singh, S.~J. Hwang, and N.~Johnston,
\newblock ``Variational image compression with a scale hyperprior,''
\newblock {\em arXiv preprint arXiv:1802.01436}, 2018.

\bibitem{theis2017lossy}
L.~Theis, W.~Shi, A.~Cunningham, and F.~Husz{\'a}r,
\newblock ``Lossy image compression with compressive autoencoders,''
\newblock in {\em International Conference on Learning Representations}, 2017.

\bibitem{van2016pixel}
A.~Van~Oord, N.l Kalchbrenner, and K.~Kavukcuoglu,
\newblock ``Pixel recurrent neural networks,''
\newblock in {\em International Conference on Machine Learning}, 2016, pp.
  1747--1756.

\bibitem{taubman2012jpeg2000}
D.~Taubman and M.~Marcellin,
\newblock {\em JPEG2000 image compression fundamentals, standards and practice:
  image compression fundamentals, standards and practice}, vol. 642,
\newblock Springer Science \& Business Media, 2012.

\bibitem{CLIC2018}
``{Challenge on Learned Image Compression},'' \url{http://compression.cc/}.

\bibitem{kodak}
``{Kodak PhotoCD dataset},'' \url{http://r0k.us/graphics/ kodak/}.

\end{thebibliography}
